\documentclass[11pt,a4paper]{article}
\usepackage{graphicx}
\usepackage{amsmath}
\usepackage{cite}
\usepackage{amsfonts}
\usepackage{amssymb}
\newtheorem{theorem}{Theorem}

\newtheorem{definition}{Definition}

\newtheorem{remark}{Remark}

\newcommand{\onetom}{1,\cdots,m}

\begin{document}

\title{Chaos synchronization in networks of coupled maps with time-varying
topologies}

\author{Wenlian Lu\thanks{Lab. of Mathematics
for Nonlinear Sciences, School of Mathematical Sciences, Fudan
University, 200433, Shanghai, China. \texttt{wenlian.lu@gmail.com}} 
\and
Fatihcan M. Atay\thanks{Max Planck Institute for Mathematics in the Sciences,
Inselstr.~22, 04103 Leipzig, Germany. \texttt{atay@member.ams.org}}
\and 
J\"urgen Jost\thanks{Max Planck Institute for Mathematics in the Sciences,
Inselstr.~22, 04103 Leipzig, Germany. \texttt{jjost@mis.mpg.de}}}


\date{Preprint. Final version in 
\emph{Eur.~Phys.~J.~B} \textbf{63}:399--406, 2008}

\maketitle

\abstract{Complexity of dynamical networks can arise  not only from
the complexity of the topological structure but also from the time
evolution of the topology. In this paper, we study the synchronous
motion of coupled maps in time-varying complex networks both
analytically and numerically. The temporal variation is rather
general and formalized as being driven by a metric dynamical system.
Four network models are discussed in detail in which the
interconnections between vertices vary through time randomly. These
models are 1) i.i.d. sequences of random graphs with fixed wiring
probability, 2) groups of graphs with random switches between the
individual graphs, 3) graphs with temporary random failures of
nodes, and 4) the meet-for-dinner model where the vertices are
randomly grouped. We show that the temporal variation and randomness
of the connection topology can enhance synchronizability in many
cases; however, there are also instances where they reduce
synchronizability. In analytical terms, the Hajnal diameter of the
coupling matrix sequence  is presented as a measure for the
synchronizability of the graph topology. In topological terms, the
decisive criterion for synchronization of coupled chaotic maps is
that the union of the time-varying graphs contains a spanning tree.

     } 
%

\section{Introduction}
Synchronization of coupled maps in networks is presently an active
research topic \cite{syn_cml}. It represents a mathematical
framework that on the one hand can elucidate -- desired or
undesired -- synchronization phenomena in diverse applications. On
the other hand, the synchronization paradigm is formulated in such
a manner that powerful mathematical techniques from dynamical
systems and graph theory can be utilized. A standard version of
the network of coupled maps, coming from the well-known coupled
map lattices (CML) \cite{Kan1}, can be formalized as follows:
\begin{equation}
x^{i}(t+1)=f(x^{i}(t))+\sum\limits_{j=1}^{m}L_{ij}f(x^{j}(t)),
i=\onetom,\label{cml-comp}
\end{equation}
where $t\in\mathbb Z^{+}=\{0,1,2,\cdots,\}$, $x^{i}(t)$ is the
state variable of vertex $i$, $f:\mathbb{R}\rightarrow \mathbb{R}$
is a differentiable map, and $L=[L_{ij}]_{i,j=1}^{m}\in
\mathbb{R}^{m\times m}$ is the diffusion matrix, which is
determined by the topological structure of the network and
satisfies $L_{ij}\ge 0$ for all $i\ne j$ and
$\sum_{j=1}^{m}L_{ij}=0$ for all $i=\onetom$. Let $x=[x^{1},x^{2},
\dots,x^{m}]^{\top}\in \mathbb{R}^{m}$,
$F(x)=[f(x^{1}),f(x^{2}),\dots,f(x^{m})]^{\top}$, and $G=I_{m}+L$,
where $I_{m}$ denotes the identity matrix of dimension $m$. Then,
Eq. (\ref{cml-comp}) can be rewritten in the matrix form:
\begin{eqnarray}
x(t+1)=G F(x(t))\label{CMLs}
\end{eqnarray}
where $G=[G_{ij}]_{i,j=1}^{m}\in \mathbb{R}^{m\times m}$ denotes the
coupling and satisfies $G_{ij}\ge 0$ for $i\ne j$ and
$\sum_{j=1}^{m}G_{ij}=1$ for all $i=\onetom$. Thus, if $G_{ii}\ge 0$
holds for all $i=\onetom$, then $G$ is a stochastic matrix.

This dynamical system formulation contains two aspects. One of them
is the reaction dynamics at each vertex of the network. The other is
the coupling structure, that is, whether and how strongly, the
dynamics at one vertex is directly influenced by the states of the
other vertices. This influence can be described by notions of graph
theory. Hence, the coupling matrix $G$ corresponds to a graph
$\Gamma=[{\mathcal V},{\mathcal E}]$, where ${\mathcal
V}=\{1,2,\cdots,m\}$ denotes the vertex set and ${\mathcal
E}=\{e_{ij}\}$ denotes the edge set such that there exists a
directed edge from vertex $j$ to vertex $i$ if and only if
$G_{ij}>0$.

Synchronous dynamics in complex networks have recently attracted
increasing attention \cite{syn_cml,Pik,Mor,Shi,Jost,Wu}. Linear
stability analysis was used and transverse Lyapunov exponents were
introduced to analyze the influence of the topological structure
of networks \cite{Jost}. Ref. \cite{Wu} has related the ability to
synchronize chaotic maps to the existence of a spanning
tree in the corresponding graph. However, synchronization analysis
has so far been mostly limited to autonomous systems, where the
interactions between the state components are static. In
\cite{Jost}, a generalized criterion guaranteeing synchronization
in the model (\ref{CMLs}) is proposed as follows:
\begin{eqnarray}
\log|\lambda_{1}|+\mu<0,\label{static}
\end{eqnarray}
where $\mu$ is the Lyapunov exponent of the uncoupled system
$s(t+1)=f(s(t))$ and $\lambda_{1}$ is the eigenvalue of the
coupling matrix $G$ with the second largest modulus, noting that
the largest eigenvalue has a modulus of $1$.

Many real-world applications from the social, natural, and
engineering disciplines  include a temporal variation of topology of
the network. In communication networks, for example, one must
consider dynamical networks of moving agents. In this case, some of
the existing connections can fail simply due to occurrence of an
obstacle between agents \cite{Sab}. Also, some new connections may
be created when one agent enters the effective region of other
agents \cite{Hat,Vic,More}. Furthermore, this temporal variation of
topology involves randomness. In \cite{Sab,Hat,Vic}, consensus in
multi-agent networks was considered where the state of each vertex
is updated according to the states of its connected neighbors with
switching connecting topologies. The consensus protocol of
multi-agent dynamical networks can generally be formalized in
discrete-time form as
\begin{eqnarray}
x^{i}(t+1)=\sum\limits_{j=1}^{m}G_{ij}(t)x^{j}(t),~i=\onetom,\label{multi}
\end{eqnarray}
where $[G_{ij}(t)]_{i,j=1}^{m}$, $t\in\mathbb Z^{+}$, are stochastic
matrices. It was proved in Ref. \cite{More} that the connectivity of
the switching graphs plays a key role in the consensus dynamics of
multi-agent networks with switching topologies. Some papers from the
recent literature \cite{Lv} studied synchronization of
continuous-time dynamical networks with time-varying topologies;
however, the time-varying couplings were specific, with either
symmetry, node balance, or fixed time average.

In this paper, we study the local complete synchronization of
networks of coupled maps with time-varying couplings:
\begin{eqnarray}
x(t+1)=G(\theta^{(t)}\omega)F(x(t)).\label{RDS}
\end{eqnarray}
Here, $\theta^{(t)}\cdot$ represents a metric dynamical
system
$\{\Omega,{\mathcal F},P,\theta^{(t)}\}$, where $\Omega$ is the
state space, $\mathcal F$ is the $\sigma$-algebra, $P$ is the
probability measure, and $\theta^{(t)}$ is the semi-flow satisfying
$\theta^{(t+s)}=\theta^{(t)}\circ\theta^{(s)}$,
where $\theta^{(0)}$ is the identity map,
$G(\theta^{(t)}\omega)=[G_{ij}(\theta^{(t)}\omega)]_{i,j=1}^{m}
\in\mathbb R^{m\times m}$ denotes the coupling matrix at time $t$
and is measurable on $(\Omega,\mathcal F)$, and
$F(x)=[f(x_{1}),\cdots,f(x_{n})]^{\top}$ is a differentiable
function.

Thus, Eq. (\ref{RDS}) is a random dynamical system. For more details
on random dynamical systems, we refer to the textbooks \cite{Arn}.
This form of time-varying coupling is rather general and includes
the deterministic case, where $G(\cdot)$ can be regarded as a known
function of time $t$, as well as the stochastic case, where
$G(\cdot)$ can be regarded as being enforced by a stochastic process
$\{\xi^{t}\}_{t\in\mathbb Z^{+}}$, namely, $G(\xi^{t})$.

Accordingly, we denote time varying graphs by
$\{\Gamma(\theta^{(t)}\omega)\}_{t\in {\mathbb Z}^{+}}$. Define
$\Gamma(\theta^{(t)}\omega)=[{\mathcal V},{\mathcal
E}(\theta^{(t)}\omega)]$, where ${\mathcal V}=\{1,2,\cdots,m\}$
denotes the fixed vertex set and
 ${\mathcal
E}(\theta^{(t)}\omega)=\{e_{ij}(\theta^{(t)}\omega)\}$ denotes the
edge set of the graph at time $t$, i.e., edge
$e_{ij}(\theta^{(t)}\omega)$ exists if and only if
$G_{ij}(\theta^{(t)}\omega)>0$. So, the coupling matrix
$G(\theta^{(t)}\omega)$ might be a function of the coupling graph
topology.

Local complete synchronization (synchronization for short) is
defined in the sense that the differences between states of vertices
of the coupled dynamical system (\ref{RDS}) converge to zero
whenever the initial state of each vertex is picked sufficiently
near the attractor of the uncoupled system and their differences are
sufficiently small, i.e., 
\begin{eqnarray}
\lim_{t\rightarrow\infty}\|x^{i}(t)-x^{j}(t)\|=0,~i,j=\onetom.
\end{eqnarray}
For a more geometric definition, 
suppose that the uncoupled system $s(t+1)=f(s(t))$
possesses an attractor (see Ref. \cite{Ash} for details), which we
denote by $A$. Define
\begin{eqnarray*}
{\mathcal S}=\big\{[x^{1},x^{2},\cdots,x^{m}]^{\top}\in\mathbb
R^{m}:~x^{i}=x^{j},~i,j=\onetom\big\}
\end{eqnarray*}
which is an invariant subspace of Eq. (\ref{RDS}). Let 
 $A^m$ denote the Cartesian
product $A\times\cdots\times A$ ($m$ times). We define the
synchronization manifold by the set $\mathcal A=\mathcal{S}\cap
A^m=\{[x,\cdots ,x]: x \in A\}$. In this sense, synchronization is
equivalent to the stability of $\mathcal A$.

The purpose of this paper is to study the synchronization of the
coupled map network (\ref{RDS}) with time-varying topology. Here,
the topology is generally supposed to be driven by a metric
dynamical system and the coupled network can be regarded as a random
dynamical system. We present sufficient conditions guaranteeing
synchronization. Furthermore, we show that the property that
the union of the time-varying graphs contains a spanning tree is
very important for the network's ability to synchronize chaotic
maps. Additionally, we present several time-varying network models
and study the synchronization of coupled maps on these dynamical
networks. The topological structures of these models vary in time
and include randomness. Generally, the collections of
interconnections in these networks can be regarded as Markov chains.
Besides illustrating the theoretical results, we also focus on the
variation of synchronizability of each model, which is quantitatively
measured with respect to several parameters in the model. As we
show, temporal variation and randomness can enhance synchronization
in some cases. Further examples indicate that the communication
between vertices in the dynamical networks might play an important
role in synchronizability.

\section{Theoretical analysis}
In this section, we present theoretical results on synchronization
of coupled map networks with time-varying couplings. The
mathematical results have been proven in detail in our companion
papers \cite{Lu,Lu1}. Our main tool to investigate the synchronous
motion of the coupled system (\ref{RDS}) is the Hajnal diameter,
which was first introduced in Ref. \cite{Haj} to describe the
compression rate of a stochastic matrix and is defined as follows:
\begin{definition}
\label{def2.1} For a matrix $G$ with row vectors
$g_{1},\cdots,g_{m}$ and a vector norm $\|\cdot\|$ in
$\mathbb{R}^{m}$, the Hajnal diameter of $G$ is defined as
\begin{eqnarray}
{\rm diam}(G)=\max\limits_{i,j}\|g_{i}-g_{j}\|.
\end{eqnarray}
\end{definition}
From this definition, synchronization of the coupled system
(\ref{RDS}) can equivalently be stated as
\begin{eqnarray}
\lim_{t\rightarrow\infty}{\rm
diam}([x_{1}(t),\cdots,x_{m}(t)]^{\top})=0.
\end{eqnarray}
We can extend this concept to  matrix sequences driven by a
dynamical system:  ${\mathcal
G}(\omega)=\{G(\theta^{(t)}\omega)\}_{t\ge 0}:\Omega\rightarrow
2^{\mathbb R^{m,m}}$ for any $\omega\in\Omega$, where $2^{\mathbb
R^{m,m}}$ denotes the set composed of all subsets of $\mathbb
R^{m,m}$. For a matrix sequence ${\mathcal G}$, its Hajnal
diameter at initial data $\omega\in\Omega$ is defined by
\begin{eqnarray}
{\rm diam}(\mathcal
G(\omega))=\overline{\lim\limits_{t\rightarrow\infty}} \big\{{\rm
diam}\bigg[\prod\limits_{k=0}^{t-1}G(\theta^{(k)}\omega)\bigg]\big\}^{\frac{1}{t}},
\end{eqnarray}
where $\prod$ denotes the left matrix product:
$\prod_{k=1}^{n}A_{k}=A_{n}\times A_{n-1}\times\cdots\times
A_{1}$.  One can see that ${\rm diam}(\mathcal G(\omega))<0$ implies
that the differences between rows of the infinite matrix product
$\prod_{t=0}^{\infty}G(\theta^{(t)}\omega)$ converge to zero as
$t$ goes to infinity.

Let $s(t)$ be the synchronized state solution satisfying
$s(t+1)=f(s(t))$ for all $t\ge 0$. Let $\delta x(t)=x(t)-s(t)$.
Linearizing the system (\ref{RDS}) about $s(t)$ gives
\begin{eqnarray}
\delta x(t+1)=f'(s(t))G(\theta^{(t)}\omega)\delta x(t).\label{var}
\end{eqnarray}
Note that
\begin{eqnarray}
&&{\rm
diam}\bigg[\prod\limits_{k=0}^{t-1}G(\theta^{(k)}\omega)f'(f^{(k)}(s_{0}))\bigg]\nonumber\\
&&={\rm
diam}\bigg[\prod\limits_{k=0}^{t-1}G(\theta^{(k)}\omega)\bigg]\bigg|\prod\limits_{l=0}^{t-1}
f'(f^{(l)}(s_{0}))\bigg|.
\end{eqnarray}
Then, the Hajnal diameter of the variational system (\ref{var})
equals to ${\rm diam}({\mathcal G}(\omega))e^{\mu}$, where $\mu$
denotes the maximum Lyapunov exponent of the attractor $A$ of the
uncoupled system,
\begin{eqnarray}
\mu=\max\limits_{s_{0}\in
A}\overline{\lim\limits_{t\rightarrow\infty}}\frac{1}{t}\sum\limits_{k=0}^{t-1}\log|f^{'}(f^{(k)}(s_{0}))|.
\end{eqnarray}
This leads to the following condition
\begin{eqnarray}
{\rm diam}(\mathcal G(\omega))e^{\mu}<1,\label{thm2}
\end{eqnarray}
which guarantees that the variable vector $x(t)$ can be
synchronized by picking the initial data of $\theta^{(t)}\cdot$
as $\omega$.

Similar to the case of static network topology, we can extend the
transverse Lyapunov exponent for the matrix sequence $\mathcal G$
in direction $v\in\mathbb R^{m}$ as:
\begin{eqnarray}
\sigma(\mathcal
G,\omega,v)=\overline{\lim\limits_{t\rightarrow\infty}}\frac{1}{t}\log
\bigg\|\prod\limits_{k=0}^{t-1}G(\theta^{(k)}\omega)v\bigg\|
\end{eqnarray}
Along the synchronization direction $e_{0}=[1,1,\cdots,1]^{\top}$,\\
one has $\sigma(\mathcal G,\omega,e_{0})=0$ since $G(\cdot)$ has a common
row sum of unity. Let
$0=\sigma_{0}\ge\sigma_{1}\ge\sigma_{2}\ge\cdots\ge\sigma_{m}$ be
the Lyapunov exponents for the initial condition $\omega$, counted
with multiplicities. We have $\sigma_{1}(\omega)=\log{\rm
diam}(\mathcal G(\omega))$ according to lemma 2.7 in Ref. \cite{Lu}.
Then, the condition (\ref{thm2}) can be rewritten as
\begin{eqnarray}
\sigma_{1}+\mu<0.\label{prop1}
\end{eqnarray}
If (\ref{prop1}) is satisfied, then the coupled system (\ref{RDS})
can synchronize.

\begin{remark}
By Proposition 4.4 in Ref. \cite{Lu}, one can
see that the criterion (\ref{static}) for static networks is a
direct consequence of the criterion (\ref{prop1}).
\end{remark}
We apply the above results to the case 
where the time-varying coupling is induced by a
homogeneous Markov chain $\{\sigma^{t}\}_{t\in\mathbb Z^{+}}$
defined on a finite state space with an irreducible transition
probability matrix. Also, a homogeneous Markov chain can be
regarded as a dynamical system $(\Omega,\mathcal
F,P_{\pi},\theta^{(t)}\cdot)$ as described in the appendix. We now
consider a coupled map network with Markov jump topologies:
\begin{eqnarray}
x^{i}(t+1)=\sum\limits_{j=1}^{m}G_{ij}(\sigma^{t})f(x^{j}(t)),~i=\onetom\label{Markovcmls}
\end{eqnarray}
or in matrix form:
\begin{eqnarray}
x(t+1)=G(\sigma^{t})F(x(t)).\label{Markovmatrix}
\end{eqnarray}
Results in Ref. \cite{Lu1} indicate that in this case, $\log{\rm
diam}(\mathcal G(\omega))=\sigma_{1}(\omega)$ exists and is a
non-random number for almost every $\omega\in\Omega$. Hence for
simplicity we can write ${\rm diam}(\mathcal G(\omega))$ as ${\rm
diam}(\mathcal G)$ and $\sigma_{1}(\omega)$ as $\sigma_{1}$. 
From (\ref{thm2}), one can obtain the criterion for
synchronization of coupled maps (\ref{Markovmatrix}) as
\begin{eqnarray}
\log{\rm diam}(\mathcal G)+\mu<0.\label{thm3}
\end{eqnarray}
According to the equivalence, we can rewrite the condition
(\ref{thm3}) as the inequality (\ref{prop1}).
From the criteria (\ref{thm3})--(\ref{prop1}), the Hajnal diameter
${\rm diam}(\mathcal G)$, or equivalently, $\sigma_{1}$, can be used
to measure the synchronizability of a Markov jump graph topology
process. The question then arises under what conditions this graph
process can synchronize some chaotic dynamics, i.e., when does it
hold that ${\rm diam}(\mathcal G)<1$. The following result comes
from Theorem 4.2 in Ref. \cite{Lu} and the theory of Markov chains
\cite{Fang}.

\begin{theorem}\label{thm2.2.3}
Suppose that $G(\cdot)$ has all diagonal elements positive and the
transition probability matrix $T$ is irreducible. Then, ${\rm
diam}(\mathcal G)<1$ if and only if the graph union
$\bigcup_{i\in\underline{N}}\Gamma(i)$ possesses a spanning tree.
\end{theorem}
For the detailed proof, we refer to Ref. \cite{Lu1}. This theorem
shows that there exist cases when a Markov jump graph process can
synchronize a chaotic map (with $\mu>0$) even though at each instant
the network may be disconnected, as long as the union graph has a
spanning tree.

\section{Applications}

In the following, we will study the synchronous dynamics in four
time-varying graph process models. In each model, the number of
vertices is constant in time but the interconnections between
vertices vary, and the variation of interactions can be regarded as
a Markov chain. We expect, on the one hand, to illustrate the
theoretical results of the previous section, and on the
other hand, to numerically analyze the synchronizability as computed
by the largest nonzero Lyapunov exponent $\sigma_{1}$, by observing
the variations of $\sigma_{1}$ with respect to several parameters in
the models.

The map $f$ is chosen here as the logistic map: $f=a x(1-x)$. We
take the parameter $a=3.90$ throughout this section (hence with the
Lyapunov exponent $\mu\approx 0.5$). Thus, we can focus on the
influence of the time-varying coupling on synchronous motions by
fixing the parameter of the coupled map, which fixes the Lyapunov
exponent $\mu$ of the uncoupled system. (Note that the theoretical
results presented above do not depend on this particular choice of
chaotic dynamics.)

We realize the coupled networks via two types of coupling
configurations. The first system is the coupled map lattice via a
time varying graph process:
\begin{equation}
x^{i}(t+1)=
\left\{\begin{array}{l}f(x^{i}(t))
+ \displaystyle{\frac{\epsilon}{k_{i}(t)}}\sum\limits_{j=1}^{m}A_{ij}(t)[f(x^{j}(t)-f(x^{i}(t)],~~~\text{if }k_{i}(t)>0,\\
f(x^{i}(t)),~~~~~~~~~~~~~~~~~~~~~~~~~~~~~~~~~~~~~~~~~~~~~~~~~~\text{if }k_{i}(t)=0,\end{array}\right.\label{eg1}
\end{equation}
where $i=\onetom$, $\epsilon\ge 0$ is the coupling strength,
$A(t)$ denotes the adjacency matrix of the graph at time $t$, and
$k_{i}(t)=\sum_{j\ne i}A_{ij}(t)$ denotes the (in-)degree of vertex $i$
at time $t$.
Synchronization is measured by the time average of the variance of
the states over the network:
\begin{eqnarray*}
K=\bigg\langle
\frac{1}{m-1}\sum_{i=1}^{m}[x^{i}(t)-\bar{x}(t)]^{2}\bigg\rangle,
\end{eqnarray*}
where $\bar{x}=(1/m)\sum_{i=1}^{m}x^{i}(t)$ and
$\langle\cdot\rangle$ denotes the time average. One can regard $K$
as a function of the coupling strength $\epsilon$. Let
$[G(t)]_{ij}=\delta_{ij}(1-\epsilon)+\epsilon
k_{i}^{-1}(t)[A(t)]_{ij}$ if $k_{i}(t)>0$ and
$[G(t)]_{ij}=\delta_{ij}$ otherwise, where $\delta_{ij}$ are
the elements of the identity matrix $I_{m}$. Then, the second largest
Lyapunov exponent $\sigma_{1}$ of the stochastic matrix series
$\{G(t)\}_{t\in{\mathcal Z}^{+}}$ is also a function of $\epsilon$.
We also define $W=\sigma_{1}+\mu$, which is the largest Lyapunov exponent of
the system (\ref{eg1}) in directions transverse to the
synchronization manifold.

The second system is the dynamical multi-agent system with the
logistic output function $f$ given above. At each vertex
$i$, the state is the average of the values $f(x^{i}(t))$ of all
its neighbors and itself, i.e.,
\begin{eqnarray}
x^{i}(t+1)={\displaystyle \frac{1}{k_{i}(t)+1}}
\bigg[\sum\limits_{j\in {\mathcal N}_{i}(t)} f(x^{j}(t))+
f(x^{i}(t))\bigg],
~~~~~i=\onetom,\label{eg2}
\end{eqnarray}
where ${\mathcal N}_{i}(t)$ denotes the neighborhood of vertex $i$
in graph $\Gamma(t)$ and $k_{i}(t)$ is the degree of the vertex $i$
at time $t$. So, $G_{ij}(t)=1/(k_{i}(t)+1)$ in the form (\ref{RDS})
if vertex $j$ is linked to vertex $i$ at time $t$; otherwise,
$G_{ij}=0$.  According to the criterion (\ref{prop1}), the quantity
$\sigma_{1}$ can be utilized to measure synchronizability of the
time varying graph process of the coupled system (\ref{eg2}). A
smaller value of $\sigma_{1}$ indicates better synchronizability.
The simulation time
length is $1000$ in all cases.

\subsection{I.i.d. random graphs}
In the independent-identical-distribution (i.i.d.) random graph, the
edge for each pair of vertices can disappear or appear randomly,
independent of time and other pairs of vertices and following an
identical distribution. This is a special case of the model
introduced in Ref. \cite{Hat}. As a realization in the present
paper, at each time $t$, $\Gamma(t)$ is a $p$-random graph following
the famous Erd{\"o}s-Renyi model \cite{Erd}: for every pair $(i,j)$,
we randomly put an edge between them with probability $p$ and the
selection is statistically independent for different times $t$ and
other pairs of vertices.

\begin{figure*}[tbp]
\includegraphics[scale=0.85]{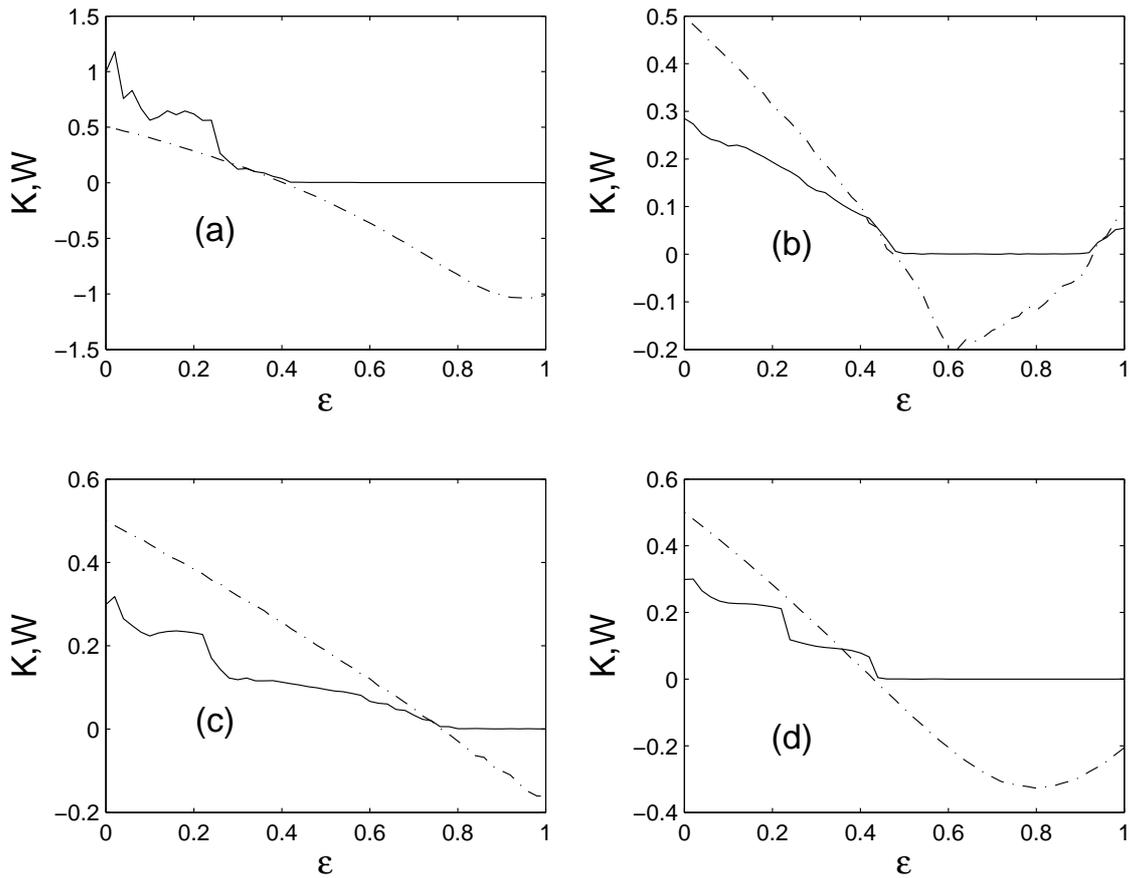}
\caption{Variation of $K$ and $W$ with respect to $\epsilon$.
Figures (a)-(d) are plotted for coupled logistic maps (\ref{eg1}).
(a) i.i.d. random network with
$200$ vertices and $p=0.1$; (b) networks with switching topologies
between $\Gamma_{3}$ and $\Gamma_{4}$ with switching probability
$p=0.5$; (c) random error model: beginning with a scale-free network introduced in
\cite{Bar} with $200$ vertices and average degree $20$, failure
occurs with probability $p=0.01$ and the recovery time $T=3$; (d)
meet-for-dinner model with $N=200$ members and subgroups of size
$n=5$ . In all cases, $K$ is shown by solid lines ($-$) and $W$ is
shown by dotted lines ($-\cdot-$).}\label{syn}
\end{figure*}

\begin{figure*}
\includegraphics[scale=0.85]{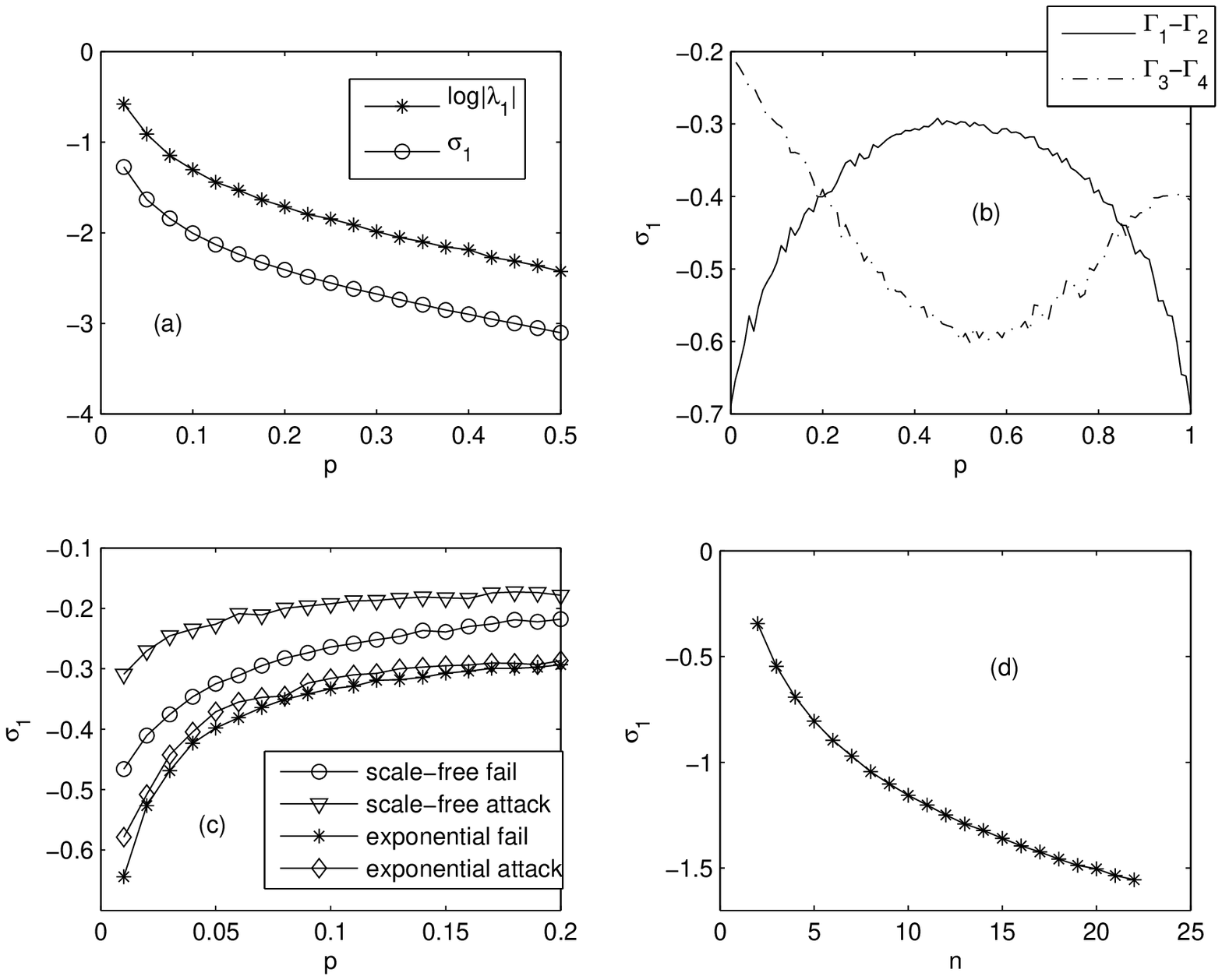}
\caption{Synchronizability of different graph processes.
Subfigures (a)-(d) are plotted for the variation of
synchronizability $\sigma_{1}$ with respect to the parameters of the
coupled network (\ref{eg2}). (a) The variation of $\sigma_{1}$ of
an i.i.d. random network with respect to the parameter $p$ and
$\log|\lambda_{1}|$  where $\lambda_{1}$ is the second largest
eigenvalue of the coupling matrix of a static random graph with
the same $p$ in the model (\ref{eg2}). The network size is 1024.
(b) The variation of $\sigma_{1}$ of a randomly switching network
with respect to $p$, for the first group
$\{\Gamma_{1},\Gamma_{2}\}$ and the second group
$\{\Gamma_{3},\Gamma_{4}\}$ of Figure \ref{switch1}. (c) The
variation of $\sigma_{1}$ of malfunction-and-recovery networks
with respect to malfunction fraction $p$, with recovery time
$T=5$, for failure and attack on scale-free and random networks.
The initial scale-free network has size $N=1024$ and average
degree $20$, and the random network has $N=1024$ and average
degree $152$. (d) The variation of $\sigma_{1}$ of the
meet-for-dinner model with respect to $n$, in a network of size
$N=1024$.}\label{synch}
\end{figure*}

We realize the coupled map networks (\ref{eg1}) and (\ref{eg2}) in
this model. Figure \ref{syn} (a) indicates that the parameter range
of the coupling strength $\epsilon$ for which synchronization occurs
coincides with the range where $W(\epsilon)<0$. This
verifies the criterion (\ref{prop1}). From the criterion
(\ref{static}), the synchronizability measure for static networks is
$\log|\lambda_{1}|$, which has been studied in e.g. Ref.
\cite{Atay}. From figure \ref{synch} (a), we observe the variation
of $\sigma_{1}$ with respect to $p$ and compare it to the logarithm
of the second largest eigenvalue (in modulus) of the coupling matrix
of the static random graph in the coupled model (\ref{eg2}). One can
see that the synchronizability of i.i.d. random graphs increases
with increasing probability parameter $p$ and is clearly better than
a static random graph of the same size and with the same wiring
probability $p$. This implies that in a random network, temporal
variation and randomness can increase synchronizability.
Furthermore, as one would expect, synchronizability increases
with the wiring probability $p$.

\subsection{Randomly switching topologies}
\begin{figure*}[tbp]
\includegraphics[scale=0.9]{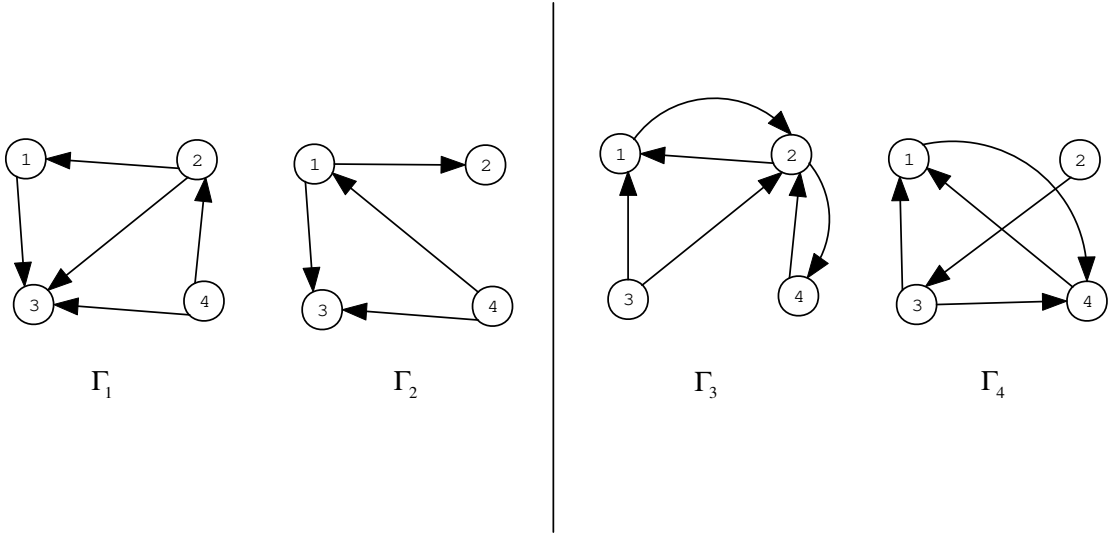}
\caption{Two groups of graphs $\{\Gamma_{1},\Gamma_{2}\}$ and
$\{\Gamma_{3},\Gamma_{4}\}$. Switch occurs in either group
randomly. Namely, at each time, with probability $p$ the graph
topology is selected as $\Gamma_{1}$ (respectively, $\Gamma_{3}$)
and with probability $1-p$ as $\Gamma_{2}$ (resp., $\Gamma_{4}$).}
\label{switch1}
\end{figure*}

Randomly switching topologies were introduced in Ref. \cite{Sab}.
That is, the graph topology at time $t$ is randomly picked from a
given finite set of topologies that follows an identical
time-independent distribution. Here, we consider two pairs of
graphs, see $\{\Gamma_{1}$, $\Gamma_{2}\}$ and $\{\Gamma_{3}$,
$\Gamma_{4}\}$ in figure \ref{switch1}. The random switch occurs
between the two graphs of each pair. The switching signal is driven
by a Bernoulli random variable $v$. For some constant $p\in(0,1)$,
if $v<p$ then the first graph in each pair is chosen as the coupling
topology; otherwise, the second graph is chosen.

From figure \ref{syn} (b), one can see that the parameter region for
which $\sigma_{1}+\mu<0$ equals to the region where $K\approx 0$,
which verifies the criterion (\ref{prop1}). From figure \ref{synch}
(b), one can see that for the graph pair
$\{\Gamma_{1},\Gamma_{2}\}$, the synchronizability of random
switching measured by $\sigma_{1}$ is worse than either of the 
individual graphs (noting that the synchronizability of each graph
can be found at the endpoints
$p=0$ and $p=1$). In contrast, for the graph pair
$\{\Gamma_{3},\Gamma_{4}\}$, the synchronizability obtained by
random switching is better than those of the individual graphs. 
That is to say, there exist
instances where temporal variation of the network topology can
increase or decrease synchronizability.

\subsection{Random errors}
In this model, we consider a network with random errors. If an error
occurs at a vertex, then  all connections of this vertex disappear.
This model is characterized by two kinds of errors. One is called
\textit{failure}, which happens to vertices following the uniform
distribution; the other is called \textit{attack}, which happens to
vertices following a selective distribution according to a certain
statistical property of the vertices. As shown in Ref. \cite{Alb},
for a class of complex networks with inhomogeneous degree
distribution (for example, the Barab\'{a}si-Albert (BA) model), the
statistics such as shortest-path diameter and clustering can have
good error tolerance if the errors occur as \textit{failure} but
they are extremely vulnerable for \textit{attacks} based on highest
degrees. As shown in Ref. \cite{Park}, the synchronizability of a
network measured by the eigenratio of the corresponding Laplacian
almost does not vary if a vertex is randomly removed but
dramatically varies by the selective removal of one vertex. In the
present paper, we realize \textit{attack} according to the
connection degree of each vertex. Namely, errors happen merely on
vertices with highest degrees. In addition, we add a recovery phase:
Every malfunctioned vertex will recover, i.e., all its connections
will appear, after a fixed time period. We denote by $p$ the
fraction of error vertices in the whole vertex set, i.e., there are
$\lfloor N\times p\rfloor$ error vertices, where
$\lfloor\cdot\rfloor$ denotes the floor function and $N$ is the size
of the whole network.

Comparing the regions of the coupling strength where $K\approx 0$
and $W<0$ in figure \ref{syn} (c), one can similarly see that the
inequality (\ref{prop1}) can precisely predict synchronization.
Figure \ref{synch} (c) indicates the variation of the
synchronizability with error occurrence. We use two network models,
namely the
Barab\'{a}si-Albert (BA) network introduced in Ref. \cite{Bar} as a
scale-free network (which has a power-law degree distribution
$P(k)\sim k^{-\gamma}$, with $\gamma=3$ independent of the size of
the network in case of sufficiently large network sizes), and a
random network with exponential tails introduced in Ref. \cite{Erd}.
One can see that for a random
network with high degrees, owing to the homogeneity of the network,
there is no substantial difference in synchronizability whether the
malfunctioned vertices are selected randomly or in decreasing order
of connection degree. On the other hand, a drastically different
behavior is observed for the scale-free network. If the
vertices with higher degrees are attacked, the synchronizability is
much reduced compared to the case with random failure. 
 Due to the degree inhomogeneity of the BA networks, the
vertices with a high degree play a more important role in
synchronization than those with smaller degrees.

\subsection{Meet for dinner}
In the meet-for-dinner model introduced in \cite{Ren}, a group of
friends decide to meet for dinner at a particular restaurant but
fail to specify a precise time. On the afternoon of the dinner
appointment, they need to find a solution to decide on the meeting
time. A centralized solution is to have an advanced conference for
the whole group; however, if this option is unavailable, then a
decentralized solution is that one meets, one at a time, a subset
in the subgroup to collect the information of this subgroup about
their expected meeting time, and update with this information
until obtaining consensus. Here, we set up the model as follows.
The whole group has $N$ members. At each time interval, the group is
randomly divided into subgroups with $n$ members (if $N \ne 0 \mod
n$, then we put the remaining ones into the last subgroup) and
each subgraph is a complete graph. Furthermore, every division is
stochastically independent of each other.

The region of the coupling strength for synchronization coincides
with the range where $W<0$ in figure \ref{syn} (d). (The tiny region of
apparent discrepancy near $\epsilon\approx 0.45$ is an artifact of
plotting the curves with finite data points.) Interestingly, the
meet-for-dinner model can synchronize a chaotic map $f$ despite the
fact that the graph is disconnected at any time. For a static
disconnected graph, there exist several vertices whose dynamical
information never reaches the others; so, obviously, a chaotic map
cannot be synchronized by a disconnected graph. However, if the
graph topology is time-varying, despite the disconnectedness of the
network at each time, the dynamical information can reach others in
a certain time period. Therefore, in this sense, theorem
\ref{thm2.2.3} implies that in some cases, temporal variation of the
network topology can enhance synchronization. Figure \ref{synch}(d)
shows that the synchronizability of the meet-for-dinner model
increases with size $n$ of the subgroups.

\section{Conclusions}
In conclusion, we have presented an effective method based on the
extended Hajnal diameter for matrix sequences to study the
synchronization in networks of coupled maps with time varying
topologies. As shown by the sufficient criteria guaranteeing
synchronization, the Hajnal diameter of the coupling matrix sequence
can be utilized to measure network synchronizability. As shown in
Sec.3, synchronizability varies with respect to several parameters
in time-varying network models. An intuitive interpretation
is that the time-cost of communication between vertices might play a
key role for synchronization of a dynamical network. The vertices
in the i.i.d. random graphs have a higher chance to access others
than in a static random graph. Attack to a network with a power-law
degree distribution is more likely to interrupt the communication
between vertices than random failures. However, for a random network
with high average degree, attack and failure can cause almost equal
damage in communication between vertices. When the network size
increases, the indirect communication of two vertices can be
enhanced by the time-varying connection structure, which can
increase synchronizability. These phenomena imply that in some cases
time-variance and randomness can enhance synchronizability. However,
as shown in figure \ref{synch} (b), it is also possible to have
decreased synchronizability. This issue deserves further
investigation in the future.

\section*{Appendix: Homogeneous Markov chain with finite state
space} A homogeneous Markov chain with finite state space and an
irreducible probability transition matrix can be regarded as a
metric dynamical system with invariant probability
$\{\Omega,\mathcal F,P,\theta^{(t)}\cdot\}$. Its state space
$\Omega={\underline{N}}^{\mathbb Z^{+}}$ is composed of all
sequences: $\omega=\{\sigma^{t}\}_{t\ge 0}$; its Borel
$\sigma$-algebra $\mathcal F=\mathcal B^{\mathbb Z^{+}}$, where
$\mathcal B$ denotes all subsets of $\underline{N}$, has a basis of
the form $\{\sigma^{t_{1}}\in B_{1},\cdots,\sigma^{t_{r}}\in
B_{r}\}$ for some $t_{1}\le t_{2}\le\cdots\le t_{r}$ and
$B_{l}\in\mathcal B$ for all $l=1,\cdots,r$; $\theta$ denotes the
shift map, $\theta \omega=\{\sigma^{(t)}\}_{t\ge 1}$; $P_{\pi}$
denotes the probability measure induced by the unique invariant
class of the transition probability matrix $\pi$, which is given by
\begin{eqnarray} 
P_{\pi}(\sigma^{t_{1}}\in B_{1},\cdots,\sigma^{t_{r}}\in
B_{r})
=\sum\limits_{i_{l}=1:l,~i_{l}\notin \mathbb
T}^{N}\sum\limits_{i_{t_{l}}\in
B_{l}:l=1,\cdots,r}\pi_{i_{1}}t_{i_{1}i_{2}}t_{i_{2}i_{3}}\cdots
t_{i_{t_{r-1}}t_{r}},
\end{eqnarray}
where $\mathbb T=\{t_{1},\cdots,t_{r}\}$, and invariant through
$\theta^{(t)}\cdot$. Induced by different initial distributions
$\xi$, this system can have different probability measures
$P_{\xi}$, but they all are not invariant over $\theta^{(t)}\cdot$.
If the invariant probability $\pi$ is ergodic in the sense that each
$\pi_{k}>0$, then for any initial distribution $\xi$, $P_{\xi}$ is
absolutely continuous with $P_{\pi}$, i.e, $P_{\xi}\ll P_{\pi}$,
which implies that any characteristic in the $P_{\pi}$ almost sure
sense certainly holds in the almost sure sense for any $P_{\xi}$ if
$\pi$ is ergodic, or equivalently, if the transition
probability matrix $T$ is irreducible. In this paper, we only focus
on the probability measure $P_{\pi}$ and simplify $P_{\pi}$-almost
surely by ``almost surely'' unless denoted otherwise. By the
multiplicative ergodic theorem for random dynamical systems
\cite{Arn}, the multiplicative Lyapunov exponents for the infinite
matrix sequence $\prod_{t=0}^{\infty}G(\sigma^{t})$ exist and are
non-random almost surely.

\end{document}